\documentclass[10pt, conference]{IEEEtran}

\usepackage{amsmath,amssymb,amsfonts}
\usepackage{graphicx}
\usepackage{textcomp}

\usepackage{amsthm}
\usepackage{xcolor}
\usepackage[figuresright]{rotating}

\usepackage{graphicx}
\graphicspath{{/}{fig/}}

\usepackage{array}
\usepackage{textcomp}
\usepackage{xcolor}
\usepackage{multirow}
\usepackage{booktabs}

\usepackage[style=base]{caption}
\usepackage{subcaption}

\usepackage{mathtools}
\usepackage{float}
\usepackage{hyperref}
\usepackage{flushend}

\usepackage{pgfplots}
\pgfplotsset{compat=1.7}
\usepgfplotslibrary{groupplots}

\usepackage{multirow,tabularx}

\newlength\figureheight
\newlength\figurewidth
\title{\LARGE \bf
    Multi-Scale Supervised 3D U-Net for Kidneys and \\ Kidney Tumor Segmentation
}

\author{
    \IEEEauthorblockN{
        Wenshuai Zhao\textsuperscript{1},
        Dihong Jiang\textsuperscript{1},
        Jorge Pe\~{n}a Queralta\textsuperscript{2},
        Tomi Westerlund\textsuperscript{2}
    }\\
    \IEEEauthorblockA{
        \textsuperscript{1} School of Information Science and Technology, Fudan University, China \\
        \textsuperscript{2} \href{https://tiers.utu.fi}{Turku Intelligent Embedded and Robotic Systems Lab, University of Turku, Finland} \\
        Emails: \textsuperscript{1}\{wezhao, jopequ, tovewe\}@utu.fi \\
    }
}

\begin{document}

\maketitle
\thispagestyle{empty}
\pagestyle{empty}

\begin{abstract}

    Accurate segmentation of kidneys and kidney tumors is an essential step for radiomic analysis as well as developing advanced surgical planning techniques. In clinical analysis, the segmentation is currently performed by clinicians from the visual inspection images gathered through a computed tomography (CT) scan. This process is laborious and its success significantly depends on previous experience. Moreover, the uncertainty in the tumor location and heterogeneity of scans across patients increases the error rate. To tackle this issue, computer-aided segmentation based on deep learning techniques have become increasingly popular. We present a multi-scale supervised 3D U-Net, MSS U-Net, to automatically segment kidneys and kidney tumors from CT images. Our architecture combines deep supervision with exponential logarithmic loss to increase the 3D U-Net training efficiency. Furthermore, we introduce a connected-component based post processing method to enhance the performance of the overall process. This architecture shows superior performance compared to state-of-the-art works using data from KiTS19 public dataset, with the Dice coefficient of kidney and tumor up to 0.969 and 0.805 respectively. The segmentation techniques introduced in this paper have been tested in the KiTS19 challenge with its corresponding dataset.

\end{abstract}

\begin{IEEEkeywords}

    Tumor Segmentation; Kidneys and Kidney Tumor Segmentation; Multi-Scale Supervision; Exponential Logarithmic Loss; 3D U-Net;

\end{IEEEkeywords}

\IEEEpeerreviewmaketitle

\section{Introduction}

Renal cell carcinoma is one of the most common genitourinary cancers with the highest mortality rate~\cite{cairns2011renal}. An accurate segmentation of kidney and tumor based on medical images, such as images from computed tomography (CT) scans, is the cornerstone to appropriate treatment. In computer-aided therapy, the success of this step is an essential prerequisite to any other processes. The segmentation process therefore is the key for exploring the relationship between the tumor and its corresponding surgical outcome, and aids doctors in making more accurate treatment plans~\cite{gillies2016radiomics}. Nonetheless, the manual segmentation of organs or lesions has the potential to be highly time-consuming, since a radiologist may need to label out target regions in hundreds of slices for one patient. The need for more accurate automatic segmentation tools is thus evident.

\begin{figure*}
    \centering
    \fbox{\includegraphics[width=0.95\textwidth]{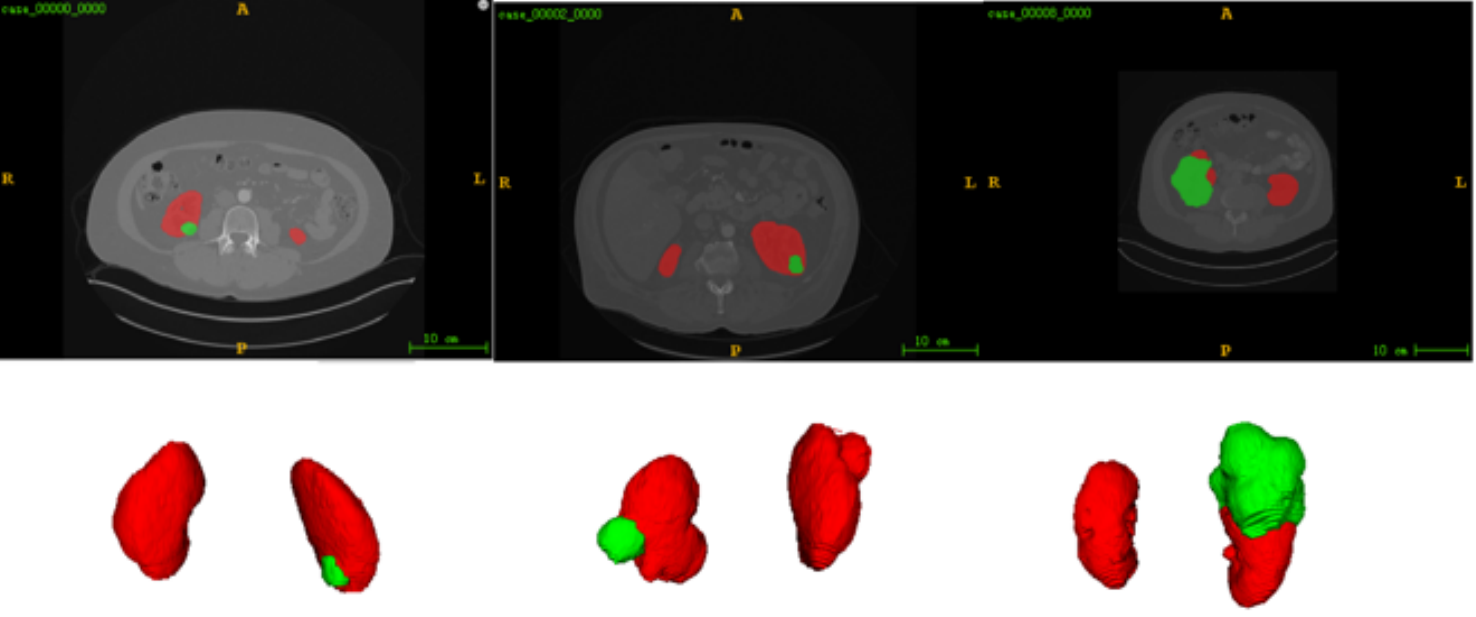}}
    \caption{Illustration of sample segmented images from three patients. The first row is the transverse plane, and the second row is the 3D reconstruction. Red voxels denote kidney while green voxels denote kidney tumor.}
    \label{fig:sample}
\end{figure*}

Multiple challenges remain in the development of automatic kidney and kidney tumor segmentation techniques from CT images within the deep learning field. Most previous works involve unsupervised training of image segmentation, including threshold-based methods, region-based methods (e.g., region growing), clustering-based methods (e.g., fuzzy C-means or Markov Random Fields), edge-detection methods, or deformable model methods~\cite{fasihi2016overview}. However, these techniques can only be applied successfully when certain conditions meet. For instance, threshold-based techniques yield the best results when the regions of interest have a significant difference in intensity with respect to the background. More recently, the application of deep artificial neural networks for medical image segmentation has gained increased momentum, in particular 3D convolutional neural networks~\cite{christ2017automatic}.

Fig.~\ref{fig:sample} shows abdominal CT images from three different patients. Because the location, size and shape of kidney and tumor vary considerable across patients, the segmentation of kidneys and kidney tumors is challenging. The major challenges can be attributed to the following considerations. First, the location of tumors may vary significantly from patient to patient. The tumor can appear anywhere inside the organs or attached to the kidneys. Trying to predict the location from experience and previous knowledge is unfeasible both from a human's and a computer's perspective. Second, the shape and size of tumors present huge diversity. Tumors in some patients can be very small on the kidneys while others can almost erode the whole kidney. Besides, their shapes might be regular, distorted, or scattered. Third, the tissue of tumors is also heterogeneous: the large amount of different subtypes of renal cell carcinoma, coupled with their heterogeneity, can bring diverse intensity attribution in CT images. Finally, the simultaneous segmentation of kidney and kidney tumor from raw full-scale CT images may lead to additional difficulties due to the co-existence of multiple labels and large background sizes.

\subsection{Background}

In recent years, deep learning methods have emerged as a promising solution for segmentation in medical imaging. Among these methods, convolutional neural network (CNN) architectures have already outperformed traditional algorithms in computer vision tasks in general~\cite{krizhevsky2012imagenet}, and in the segmentation of CT images in particular~\cite{christ2017automatic}. Fully convolutional network (FCN) architectures are a notably powerful end-to-end training segmentation solution~\cite{long2015fully}. This type of architecture is able to produce raw-scale, pixel-level labels at the images, and it is the current state-of-the-art across multiple domains. Other recent works have taken FCN as a starting point towards deeper and more complex segmentation architectures, such as the feature pyramid networks (FPN), mainly used in object detection~\cite{lin2017feature}, pyramid scene parsing networks (PSPNet), utilized in scene understanding~\cite{zhao2017pyramid}, Mark R-CNN, for object instance segmentation~\cite{he2017mask}, or DeepLab series for semantic image segmentation~\cite{krizhevsky2012imagenet}.

While there is a wide range of deep learning methods for image segmentation, medical images present significant differences from natural images. Some of the most distinct features are the following: First, medical images are relatively simple, when compared to the wide range of semantic patterns, colors and intensities in natural images. This increased homogeneity across individual images hinders the identification of patterns and regions of interest; Second, the boundary between organs, lesions or other regions of interest is fuzzy, and the images are not obtained through passive observation of the subject but instead through an active stimulus. In consequence, methods and neural network architectures employed in other types of image segmentation cannot be directly extrapolated to the medical field. When taking into account the way in which these images are obtained, the difference becomes even more significant. Medical images are often obtained through a volumetric sampling of the subject's body. This characteristic and key differential aspect can be taken as an advantage towards integrating three-dimensional neural network architectures. Among those, one of the most popular architectures to date is 3D U-Net~\cite{cciccek20163d}.

3D U-Net is one of the most well-known methods and widely used three-dimensional architecture for medical image segmentation~\cite{minaee2020image}, inspired by FCNs and encoder-decoder models. The 3D U-Net architecture has been further developed and new solutions are built on top of it, for example: Nabila et al. proposed the use of Tversky loss to enhance the performance of the attention mechanism in U-Net~\cite{abraham2019novel}, whereas Zhe et al. proposed an improved U-Net architecture in which the authors perform liver segmentation in CT images by minimizing the graph cut energy function~\cite{liu2019liver}. 

A smaller number of research have been focused on the segmentation of kidneys or kidney tumors. Yu et al. proposed a novel network architecture combining vertical patch and horizontal patch based sub-models to conduct center pixel prediction of kidney tumors~\cite{yu2019crossbar}. However, owing to the nature of the data being segmented, its training and inference processes become increasingly challenging, and this type of solution minimizes the problem with a simpler output. An ideal architecture should further extend the existing end-to-end networks for pixel-wise segmentation. In addition, the nature of three-dimension data could provide higher levels of correlation to employ. In this direction, Yang et al. combined a basic 3D FCN and a pyramid pooling module to enhance the feature extraction capability, with a network able to segment kidneys and kidney tumor at the same time~\cite{yang2018automatic}. However, the experiment is conducted on the region of interest (ROI) only, rather than on raw CT images. This significantly reduces the complexity of the segmentation task as well as the clinical practicality.

In general, we have found that most algorithms in the field of medical image segmentation take the U-Net architecture as a starting point for further developments. Fabian et al. implemented a well-tuned 3D U-Net (nnU-Net) and demonstrated its applicability and potential with top rankings in multiple medical image segmentation challenges~\cite{isensee2018nnu}. In this paper, we get inspiration from their work towards an end-to-end framework to perform segmentation of kidneys and kidney tumors simultaneously from CT images. The proposed network architecture is also developed from the original 3D U-Net architecture~\cite{cciccek20163d}. Because of a clear similarity of CT images across multiple patients, we make the assumption that the original 3D U-Net is capable of extracting sufficient features for recognition. Therefore, we do not consider the utilization of additional modules or branches behind the main 3D U-Net backbone, such as attention gates~\cite{oktay2018attention}, residual modules~\cite{milletari2016v}, or FPN~\cite{lin2017feature}, among others. Instead, we focus on optimizing the training and enhancing the performance of the original 3D U-Net architecture.

\subsection{Contribution and Structure}

In this work, we explore the potential of the 3D U-Net architecture through the combination of deep supervision and the exponential logarithmic loss. With the increased number of hyperparameters, we can train better the network and utilize our results as a baseline to analyze the performance of 3D U-Net. Thus, the main contributions of this work are the following: 
\begin{enumerate}
\item the introduction of a multi-scale supervision scheme for the 3D U-Net that tunes the network to conduct an accurate prediction from deep layers; and 
\item the utilization of exponential logarithmic loss ~\cite{wong20183d} to alleviate the class imbalance problem between foregrounds (kidney and tumor voxels) and background.
\end{enumerate}
By integrating these two approaches, we have improved the performance of the original 3D U-Net architecture. Furthermore, we have designed a connected-component based post processing method to remove disconnected voxels that are detected as evident false positives.

The remainder of this paper is organized as follows. Section 2 introduces the neural network architecture, and the different methods employed in the experiments, after which Section 3 introduces experimental results and analysis of the network's performance. In Section 4, we discuss on the different strategies being taken towards medical image segmentation. Finally, Section 5 concludes the work and outlines future research directions.

\section{Methodology}

The current approaches to medical image segmentation with deep learning can be roughly classified in two trends. First, those in which the input data to the neural networks is not the raw data but instead the regions of interest (ROIs). This naturally allows higher accuracy and performance across multiple metrics. Nonetheless, the overall performance can be significantly impacted by the uncertainty in the detection and extraction to ROIs. Second, in a more recent trend, end-to-end segmentation architectures have been introduced. In these, raw images are fed to the network and the output of the network is pixel-wise segmentation with images of the same size as the inputs. We follow this end-to-end architecture to segment both kidneys and kidney tumors simultaneously from raw volumetric CT images. This allows for direct application of our methods in clinical settings.

The CT data that is fed to a CNN consists of abdominal CT images with hundreds of slices. Each three-dimensional region is called a voxel. A typical input is 512x512x200 voxels, where 200 is the number of slices and 512x512 the resolution of each image in pixels. Owing to the large input size, it is unfeasible to fed the data into the network at once especially when graphic processing units (GPUs) are used to accelerate CNNs. In GPUs, the challenges stem from both the limited amount of memory and the required computing power in general. Therefore, we follow the recent trend in both patch-based training and inference for our network architecture.


The rest of this section describes the different steps taken for training the network and processing the data.



\subsection{Data Preprocessing}

Preprocessing raw CT images before feeding them to the network is an essential step to enable an efficient training. The first aspect to consider is the existence of unexpected materials that might appear inside patients' bodies. In particular, it is a well-known fact that metallic artifacts have a significant negative effect on the quality of CT images. The main problem with artifacts is when these create regions in the images with abnormal intensity values, either much higher or much lower than in pixels corresponding to organic tissue. Due to data-driven models that deep learning algorithms are built upon, the learning process can be significantly affected by outlier voxels corresponding to non-organic artifacts. To reduce the impact of non-organic artifacts, we perform a unified preprocessing of the complete dataset, both for training and testing data. In all images, we only consider valid the intensity range between the 0.5th and 99.5th percentiles, and clip outlier values accordingly. After preprosessing, the data is normalized with a normal foreground mean and standard deviation to improve the training of the three-dimensional network.

Another preprocessing technique necessary to appropriately train a three-dimensional network from the KiTS19 dataset is the unification of the voxel space across different image sets. This is necessary because even though the transverse plane is always formed by a constant number of pixels, the corresponding voxel size may change. Therefore, failing to unify the voxel space will lead to different data inputs representing different volumes. Such anisotropy of 3D data might diminish the advantage of using 3D convolution, and end up leading to a worse performance than 2D networks can achieve. Therefore, we chose to resample the original CT images into the same voxel space if they are not, even though the resampled images often end up being of different sizes.



\subsection{Data Augmentation}

\begin{figure*}
    \centering
    \includegraphics[width=0.95\textwidth]{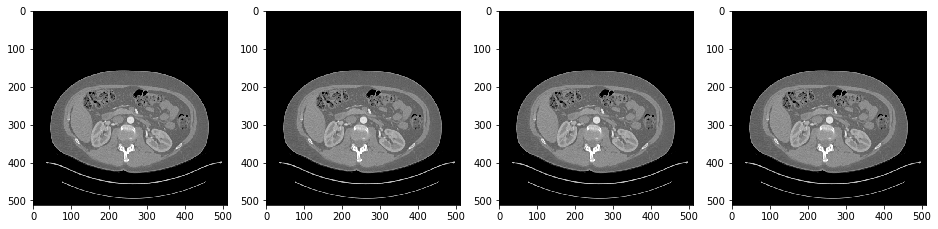} \\ 
    \includegraphics[width=0.95\textwidth]{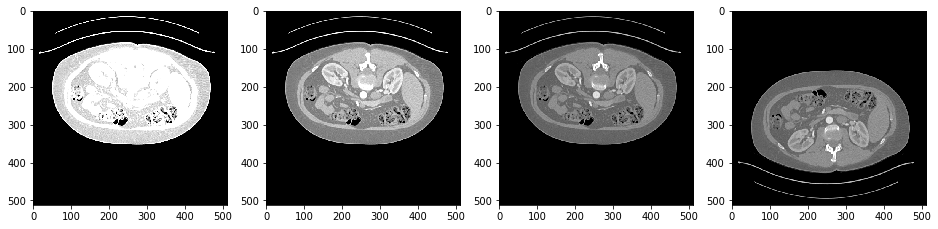} \\
    \includegraphics[width=0.95\textwidth]{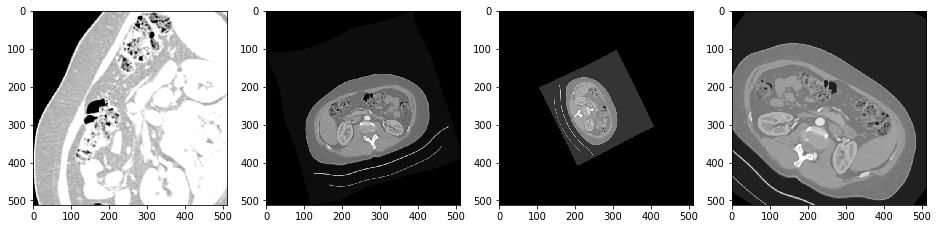} \\
    \caption{The effect of data augmentation techniques used in our method. First row is the original images; second row shows the effect of contrast augmentation and mirroring; third row shows elastic deforming, scaling, gamma correction and rotation. Of note, we show 2D images instead of the actual 3D ones for simplicity.}
    \label{fig:augmentation}
\end{figure*}

Annotating a medical image dataset can often be a lengthy and challenging task. This has so far limited the availability and size of labelled datasets. At the same time, the more complex deep learning methods become, the more data is needed to train the networks. This becomes even more critical with three dimensional networks due to the inherent increase of parameters that comes with the extra dimension. Insufficient training data would lead to overfitting and devaluate the advantage of deep learning. To solve this, a typical step is to utilize different augmentation techniques to increase the amount of available data while avoiding overfitting as much as possible. We conduct a variety of data augmentation techniques on our limited training data in order to obtain an enhanced performance of the trained network. These techniques are implemented based on the batchgenerator framework\footnote{\href{https://github.com/MIC-DKFZ/batchgenerators/}{https://github.com/MIC-DKFZ/batchgenerators/}} including random rotations, random scaling, random elastic deformations, gamma correction augmentation and mirroring. Visualization of their effect is shown in Fig.~\ref{fig:augmentation}.


\subsection{Network Architecture}

\begin{figure*}
    \centering
    \fbox{\includegraphics[width=0.95\textwidth]{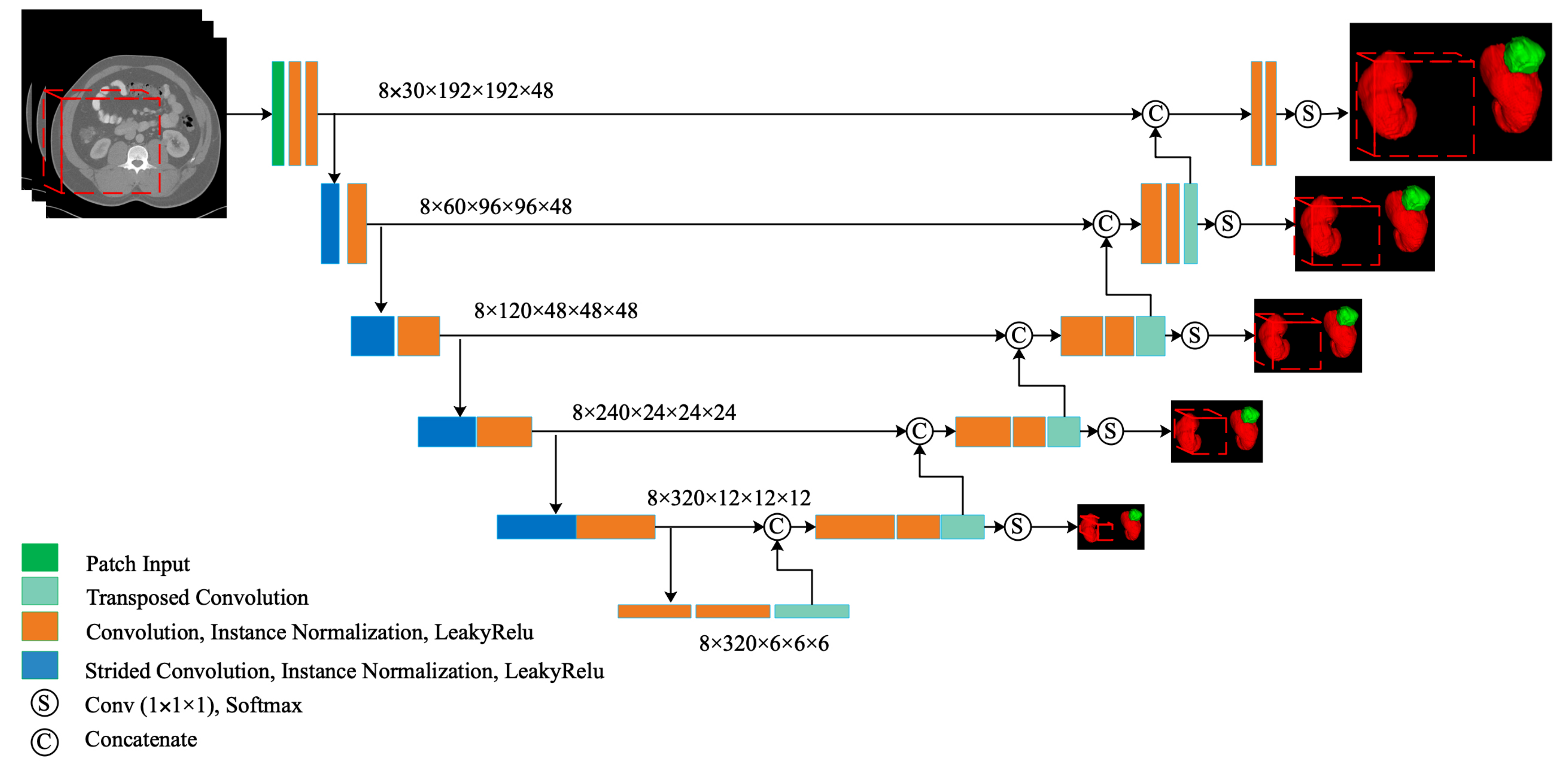}} \\ 
    \caption{The architecture of our proposed multi-scale supervised 3D U-Net. For simplicity, we use 2D icons instead of actual 3D ones, and it is best viewed in color.}
    \label{fig:architecture}
\end{figure*}

The network architecture defined in this paper has been designed taking the nnU-Net neural network framework as a starting point~\cite{isensee2018nnu}. The nnU-Net framework, unlike other recently published methods, does not add complex submodules and instead is mostly based on the original U-Net architecture. In addition to the default framework, we utilize multi-scale supervision to enhance the network's segmentation performance. The architecture of our proposed network is illustrated in Fig.~\ref{fig:architecture}, where two-dimensional images have been utilized for illustrative purposes even though the network's layers are three-dimensional. Following the main structure of 3D U-Net, the network implements decoder (left side) and encoder (right side) elements. The encoder layers are utilized to learn feature representations from the input data. The decoder is then employed for retrieving voxel locations and for determining their categories based on the semantic information extracted from encoder path. 


We adopt strided convolution instead of common pooling operation to implement downsampling, which could avoid significant negative effect on the fusion of position information. Furthermore, we replace trilinear interpolation with transposed convolution to enable adaptive upsampling. In multiple previous works, normalization is usually deployed between any two layers to obtain fixed input distribution. However, since the batch size we used is limited by the GPU memory capacity, in this work we employ instance normalization instead of batch normalization. Batch normalization naturally suits better bigger batch sizes.

The depth of the network is often decided as a trade-off between the amount of semantic features and the spatial accuracy of the segmentation. The deeper the layer, the more semantic information can be extracted. At the same time, deeper layers tend to lose location information due to the continuous decrease of resolution of the feature maps. Therefore, we have set the architecture of our network to only 6 layers, including the bottleneck, so that the deepest feature map is not smaller than 8×8×8. At the same time, to reduce the model volume, we set the basic kernel number as 30. In addition, a short connection between the encoder and decoder layers is built in the U-Net to enable the utilization at the decoder of more exact location information embedded in the encoder part.

In contrast with the original 3D U-Net architecture, we construct multi-scale supervision to encourage every layer in the decoder path to achieve exact location as well as semantic information. The motivation behind of our proposed multi-scale supervision is comprehensively depicted in the next subsection.

\subsection{Multi-scale Supervision}

In traditional deep learning based segmentation, the model outputs the probability map from the top layer, which does not fully utilize the  deeper feature maps even though they might contain more semantic information. As the top layer is upsampled from the deep layers, it is reasonable to in advance guarantee the deep layers with correct predictions. This would expected to provide a better basis to up layers in turn. Therefore, we add labels with corresponding resolution to each layer in the decoder path, and compare them with the side outputs from deep layers. With the loss calculated from different layers, more effective gradient backpropagation can be obtained and thus increase the learning efficiency. The loss function for each iteration from multi-scale supervision is given by~\eqref{eq1}:

\begin{equation}
    Loss_{t}=\displaystyle\sum_{l=1}^{N} Loss_{l}{w_{l}}
    \label{eq1}
\end{equation}

\noindent where $Loss_{t}$ denotes the total loss achieved by the multi-scale supervision, while $Loss_{l}$ refers to the loss calculated at the $l$-th layer. The total number of layers with $N$, excluding the bottleneck layer in the network. Finally, the weight of each individual layer's loss is given by $w_l$.

\subsection{Loss Function}

Cross entropy (CE) is a widely used pixel-wise loss as it computes the entropy of prediction probability and ground truth based on each pixel. However, such property might lead to severe sample imbalance since the background occupies most of the CT images.  Its definition is given by~\eqref{eq2}:

\begin{equation}
    CE = -\displaystyle\sum_{c \in\: classes} w_{c} y_{true}\log(y_{pred})
    \label{eq2}
\end{equation}

\noindent where $y_{true}$ represents the ground truth and $y_{pred}$ the predicted probability. The weight $w_c$ for each of the classes is utilized to adjust the global cross entropy value.

Another key loss function is the Dice coefficient (Dice). The Dice differs from the cross entropy in that it is especially useful for tiny target segmentation as it calculates the similarity of predicted result and ground truth regardless of the target’s relative size. Its definition for a class is given by~\eqref{eq3}:

\begin{equation}
    Dice = \dfrac{2\vert U_{pred} \cap U_{true} \vert}{\vert U_{pred} \vert + \vert U_{true} \vert}
    \label{eq3}
\end{equation}

\noindent where $U_{pred}$ and $U_{true}$ represent the segmentation results set (predictions) and the ground truth set, respectively, and $\vert U_i \vert$ denotes the cardinality of set $U_i$. The Dice is proportional to intersection of both sets, and therefore is affected by both false positives and false negatives. The higher the value of Dice, the better the segmentation effect is. Additionally, we utilize the Soft Dice ($SD$) coefficient, as it is beneficial to directly utilize the prediction probabilities and not the binary mask that the softmax layer generates. This allows us to have a cost function that is more sensitive to the learning process, and adjust the network weights more efficiently during the learning process. In order to have a function to minimize,  $1-Dice$ is often utilized as a cost function. Thus, in practice, \eqref{eq3} turns into \eqref{eq:softdice} for each class mask:

\begin{equation}
    SD_{cost} = 1 - \dfrac{2\displaystyle\sum_{pixels}y_{pred}y_{true}}{\displaystyle\sum_{pixels} y_{pred}^2 + \displaystyle\sum_{pixels} y_{true}^2}
    \label{eq:softdice}
\end{equation}

The nnU-Net implementation utilizes both cross-entropy and $1-Dice$ for training. However, in kidney and kidney tumor segmentation additional challenges arise leading us to choose a different cost function. First, the number tumor samples in the CT images is significantly smaller than the number of background and kidney samples. Second, the morphological heterogeneity of tumor voxels is significantly larger than that of kidney voxels. This imbalance of both samples and difficulty is a generic problem in medical image segmentation and would cause tendency of the network to incorrectly categorize tumor voxels. Therefore, in this paper, we modify the nnU-Net to perform exponential logarithmic loss on our Soft Dice loss to alleviate this imbalance. The cost function utilized for minimization durint the trainig is given by~\eqref{eq5}:

\begin{equation}
    SD_{ell} = (-\log{SD_{kidney}})^{0.3}\times0.4+(-\log{ SD_{tumor}})^{0.3}\times0.6
    \label{eq5}
\end{equation}

\noindent where $SD_{ell}$ represents the Soft Dice modified by exponential logarithmic loss. $SD_{kidney}$ and $SD_{tumor}$ denote the original Soft Dice calculated on kidney and tumor respectively:

\begin{equation}
    SD_{kidney} = \dfrac{2\displaystyle\sum_{\substack{kidney\\pixels}}y_{pred}y_{true}}{\displaystyle\sum_{\substack{kidney\\pixels}} y_{pred}^2 + \displaystyle\sum_{\substack{kidney\\pixels}} y_{true}^2}
\end{equation}

\begin{equation}
    SD_{tumor} = \dfrac{2\displaystyle\sum_{\substack{tumor\\pixels}}y_{pred}y_{true}}{\displaystyle\sum_{\substack{tumor\\pixels}} y_{pred}^2 + \displaystyle\sum_{\substack{tumor\\pixels}} y_{true}^2}
\end{equation}

This operation of nonlinearity could enable the network to achieve higher loss when the samples are very difficult to recognize, i.e. the prediction result is very bad, and only when the prediction is good above a certain threshold will the loss decline dramatically. In this way, the network with exponential logarithmic loss will potentially obtain more efficient gradient updates than when using linear loss functions. In addition, to further induce the network into increasing the significance of tumor samples during training, we attribute different multipliers, 0.4 and 0.6, as weights to kidney and tumor, respectively. 

Finally, we combine the Soft Dice with exponential logarithmic loss and CE as the loss function for each layer, shown in~\eqref{eq6}:

\begin{equation}
    Loss_{layer} = SD_{ell}+CE
    \label{eq6}
\end{equation}

\noindent where the $Loss_{layer}$ denotes the loss we obtain from each layer. With multi-scale supervision, we further assign different weights on different layers. As a result, from the top down, the total 5 layers excluding the bottleneck get 0.4, 0.2, 0.2, 0.1, 0.1 respectively and we attribute 0.28, 0.28 and 0.44 for the CE weights of background, kidney and tumor, respectively, to further emphasize tumor samples. Our method brings such hyperparameters combining the multi-scale supervision with exponential logarithmic loss, which can be further optimized to extract more potential out of the 3D U-Net. While there is plenty of room for optimization of these parameters, our experiments already show the benefits and an enhanced segmentation performance.

\subsection{Inference and Postprocessing}

Because network training and inference are conducted patch-wise, the accuracy of the border of patches is decreased compared with the patch center. Therefore, we adopt overlap prediction and weigh more on the center values when aggregating predictions across patches. Patches are chosen to overlap by one half of the patch size. Furthermore, we employ test-time data augmentation by mirroring testing patches along all valid axes to aggregate more predictions as well as adding Gaussian noise. Thus, for every voxel there are multiple predictions aggregated to determine its optimal value. In the center of patient data, up to 64 predictions from several overlaps and mirroring are aggregated up.

After the inference by the network, we utilise the basic knowledge to further improve the performance: most humans have two kidneys, and the kidney tumor should be attached on or embedded in kidneys. Even though this is very elementary information, it can be employed in postprocessing to remove disconnected voxels that are detected as false positive. False positive kidney voxels are removed from the output after either one or two kidney components have been found. Thus, all tumor components attached to a kidney are maintained as valid segmentation result while others are removed from the output as well.

\begin{figure*}
    \centering
    \fbox{\includegraphics[width=0.95\textwidth]{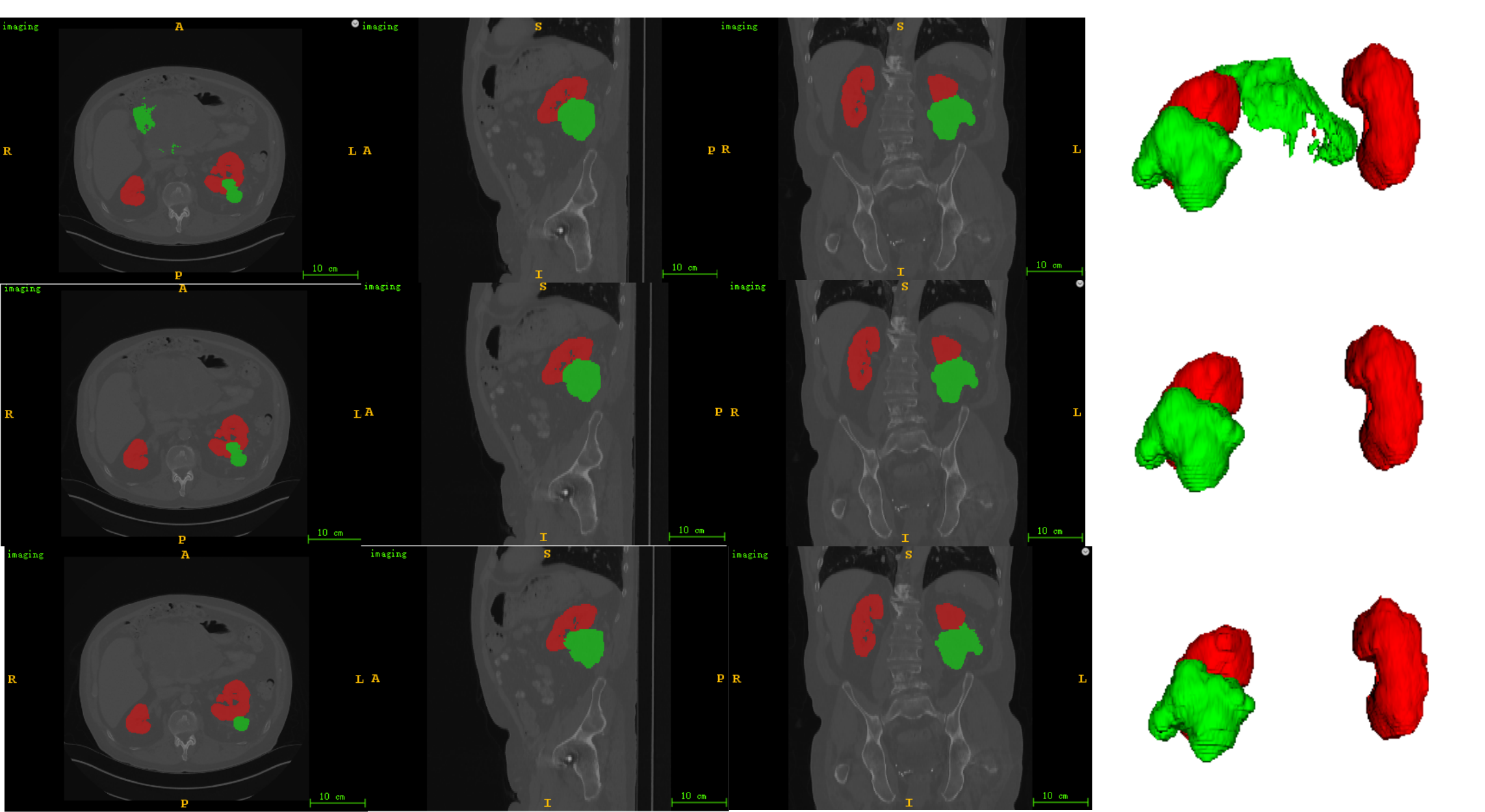}} \\ 
    \caption{The effect of our postprocessing method. The top row is original output from the network; the middle is after post processing and the bottom is the ground truth.}
    \label{fig:postprocessing}
\end{figure*}

Fig.~\ref{fig:postprocessing} shows how our postprocessing method increases the quality of the segmentation. In the top most figure, we have two kidneys and a disconnected tumor. The segmentation keeps the top two biggest kidney components as shown in the middle of the figure. The result after the post-processing effectively matches the ground truth sample. It should be accounted that if there is only one kidney, the patient might have previously received a nephrectomy.

\section{Experimental Results}

In order to evaluate the proposed three-dimensional network architecture, and to validate its usability with real data, we conduct experiments on the public KiTS19 dataset. We separate the KiTS19 dataset into the training, validation and test datasets, which we utilize to train and evaluate our model's performance with both visualization metrics and quantification metrics.

\subsection{KiTS19 Dataset}

The KiTS19 dataset contains volumetric CT scans from 210 patients. These scans are all preoperative abdominal CT imaging in the late-arterial phase, with unambiguous definition of kidney tumor voxels in the ground truth images. Scans of different patients have different properties. Therefore, there is a heterogeneity in the raw data, including the voxel size along the three plans and their affine. As diverse voxel spacings may have a significantly negative impact on the learning process of deep neural networks, we use instead the interpolated dataset in KiTS19, which interpolates the original dataset to achieve the same affine for every patient. 

The statistic properties of the dataset are shown in Table~\ref{tab1}. We randomly select 20\% of the patients as the independent test dataset. Among the rest, we partition the records into validation dataset (20\% of scans) and training dataset (80\% of scans).

\begin{table*}
    \centering
    \caption{Properties of the KiTS19 dataset}
    \label{tab1}
    \begin{tabular}{@{}ll@{}}
        \toprule
        \textbf{Property}\hspace{5cm} & \textbf{Value}           \\
        \midrule
        Number of Patients              & 210                      \\
        Modality                        & CT (late-arterial phase) \\
        Training dataset size           & 134                      \\
        Validation dataset size         & 34                       \\
        Test dataset size               & 42                       \\
        Min Patient Size in Voxels      & {[}434, 434, 69{]}       \\
        Max Patient Size in Voxels      & {[}639, 639, 182{]}      \\
        Median Patient Size in Voxels   & {[}523, 523, 116{]}      \\[+3pt]
        Affine                          & 
            $ \left(
                    \begin{array}{cccc}
                        0   &  0 & -0.7816 &  0  \\
                        0 & -0.7816 & 0 & 0 \\
                        -3 & 0 & 0 & 0 \\
                        0 & 0 & 0 & 1 \\
                    \end{array}
                \right) $ \\[+6pt]
        \bottomrule
    \end{tabular}
\end{table*}

\subsection{Implementation Details}

We utilize Adam as the network's optimizer function and set the initial learning rate to be 3×10$^{-4}$. In addition, we conduct an adaptive adjustment strategy for the learning rate during the training process. This results in the learning rate dropping by the factor of 0.2 whenever the training loss is not improved over 30 epochs. Similarly, we consider that the training is over whenever no improvements in the loss are identified over 50 epochs. The complete process is implemented in Python utilizing the PyTorch framework.

In the experiments, the training is carried out with two Nvidia Tesla 32 GB GPUs. Owing to the limited amount of the GPU memory, we adopt a patch size of 192×192×48 and set the batch size to eight. Since the training is patch-based, the patch is randomly sampled from the data loader and we each epoch is set to 250 iterations. This translates into each epoch effectively selecting 250×8 patches from the training data.

\subsection{Experiment Results}

\begin{figure}
    \centering
    \includegraphics[width=0.45\textwidth]{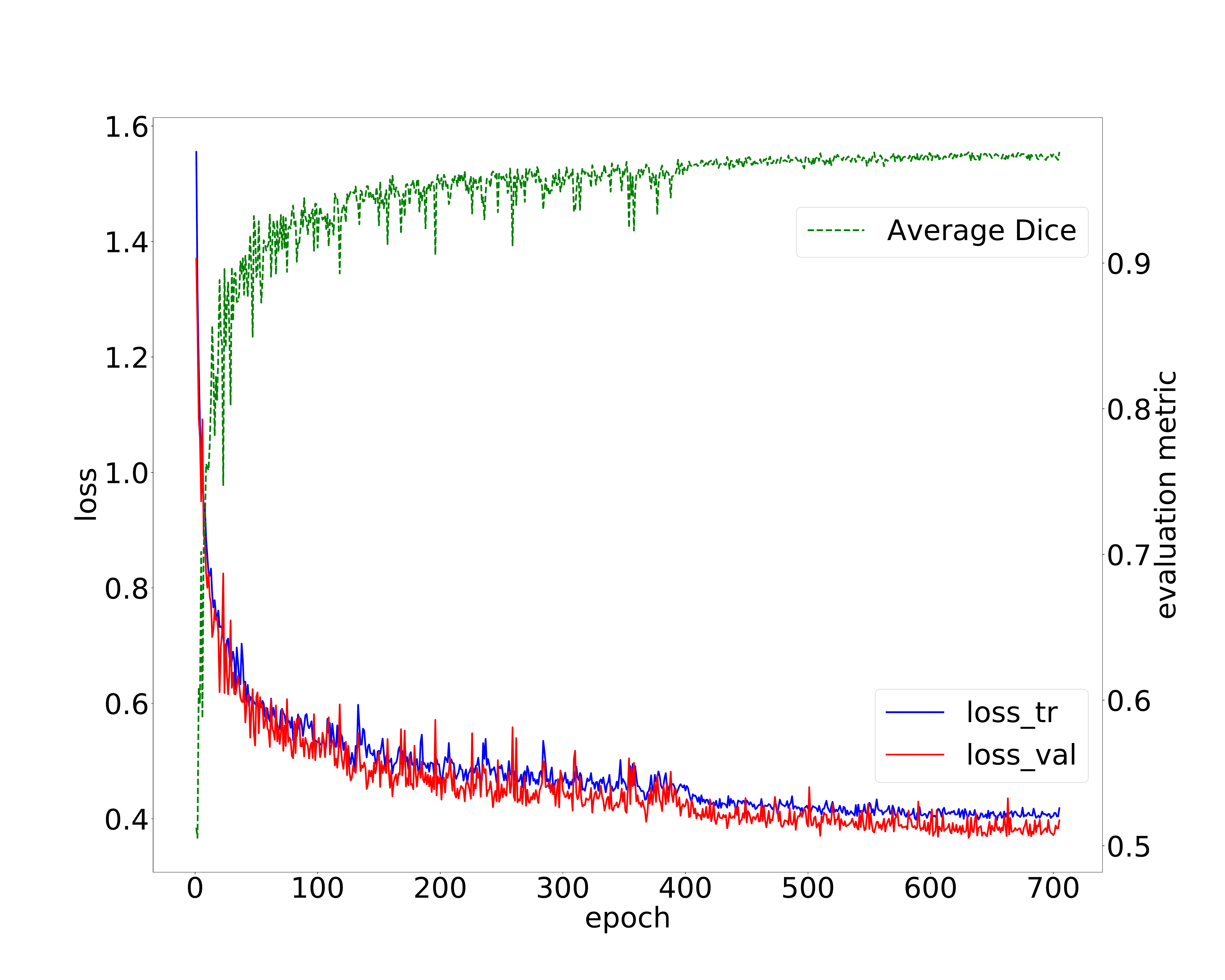}
    \caption{Loss and Dice evolution during the network training. The red and blue lines represent the validation and training loss, respectively. The green line represents the average Dice of kidneys and kidney tumor. The total training time was approximately five days.}
    \label{fig:loss}
\end{figure}

The training time until convergence was achieved with our network architecture was approximately five days. The loss variation is shown in Fig.~\ref{fig:loss}, from which we can observe that the loss decreases stably as the number of training epoch increases.

\begin{figure*}
    \centering
    \fbox{\includegraphics[width=0.95\textwidth]{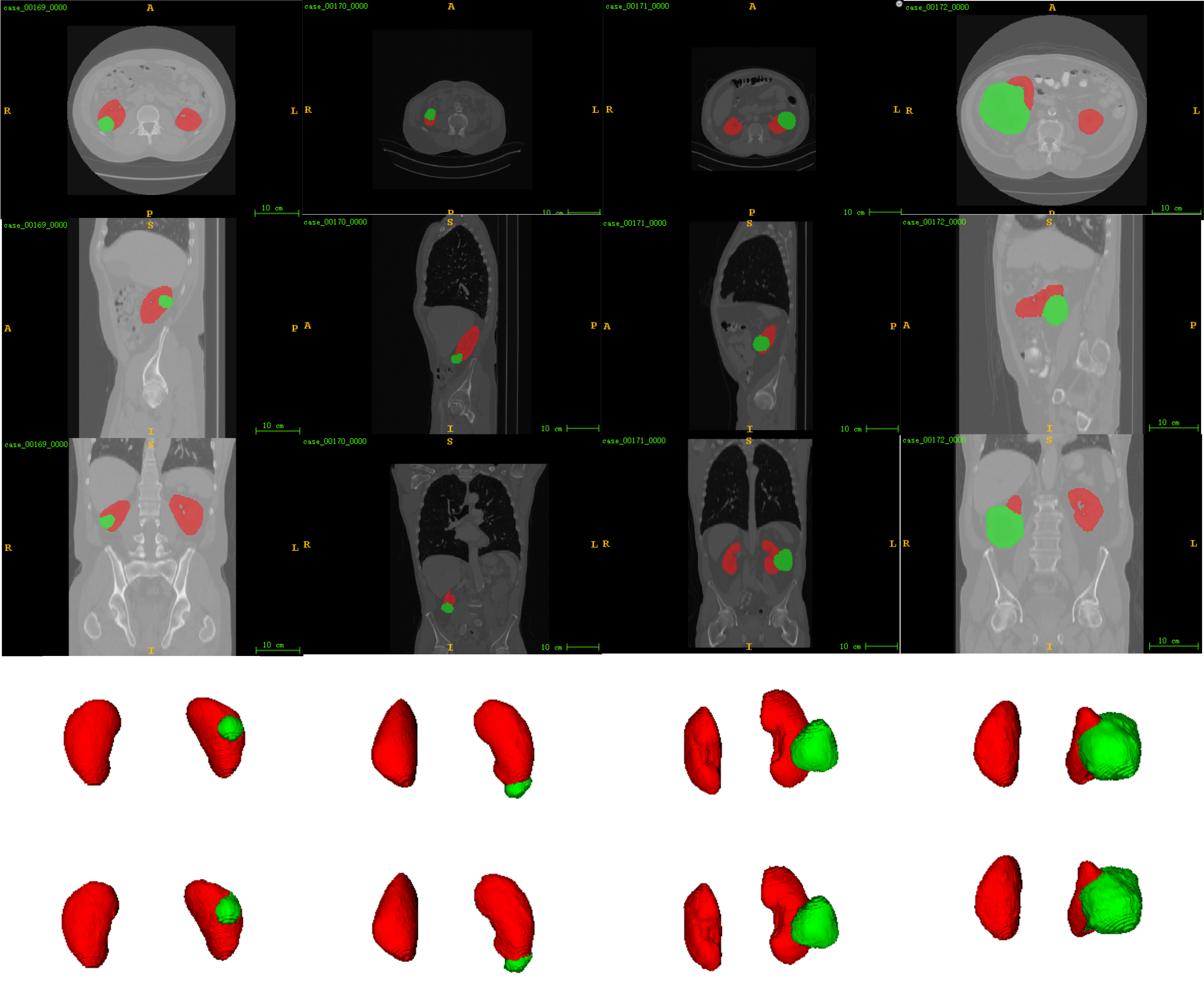}}
    \caption{Examples of segmentation results. Each column denotes one patient, and from top down, they are scan images in transverse plane, sagittal plane, coronal plane, 3D reconstruction of our predictions and the ground truth. The red mask represents kidney area, while the green mask represents tumor area.}
    \label{fig:results}
\end{figure*}

After the model is trained and convergence achieved, we test it using the independent test dataset containing 42 patient scans. Several selected samples of output segmentation results are shown in Fig.~\ref{fig:results}. Even though the location, intensity and texture of kidneys and kidney tumor can differ significantly, the predicted regions are in good agreement with the ground truth. 

\begin{figure}
    \captionsetup[subfigure]{aboveskip=-0.3\baselineskip}
    \centering
    \begin{subfigure}[t]{0.495\textwidth}
        \includegraphics[width=\textwidth]{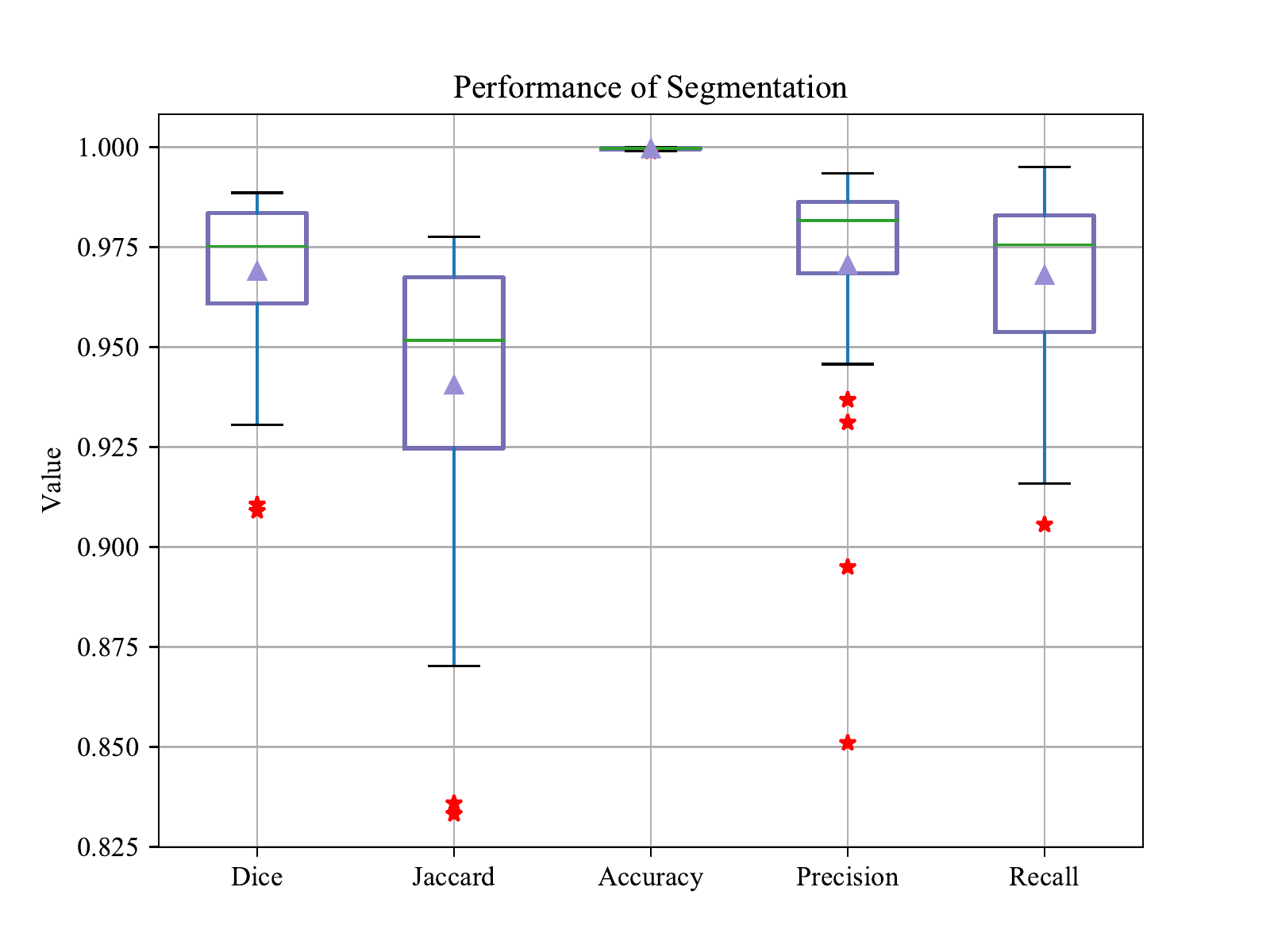} 
        \caption{Kidney segmentation}
        \label{fig:kidney}
    \end{subfigure}
    \begin{subfigure}[t]{0.495\textwidth}
        \includegraphics[width=\textwidth]{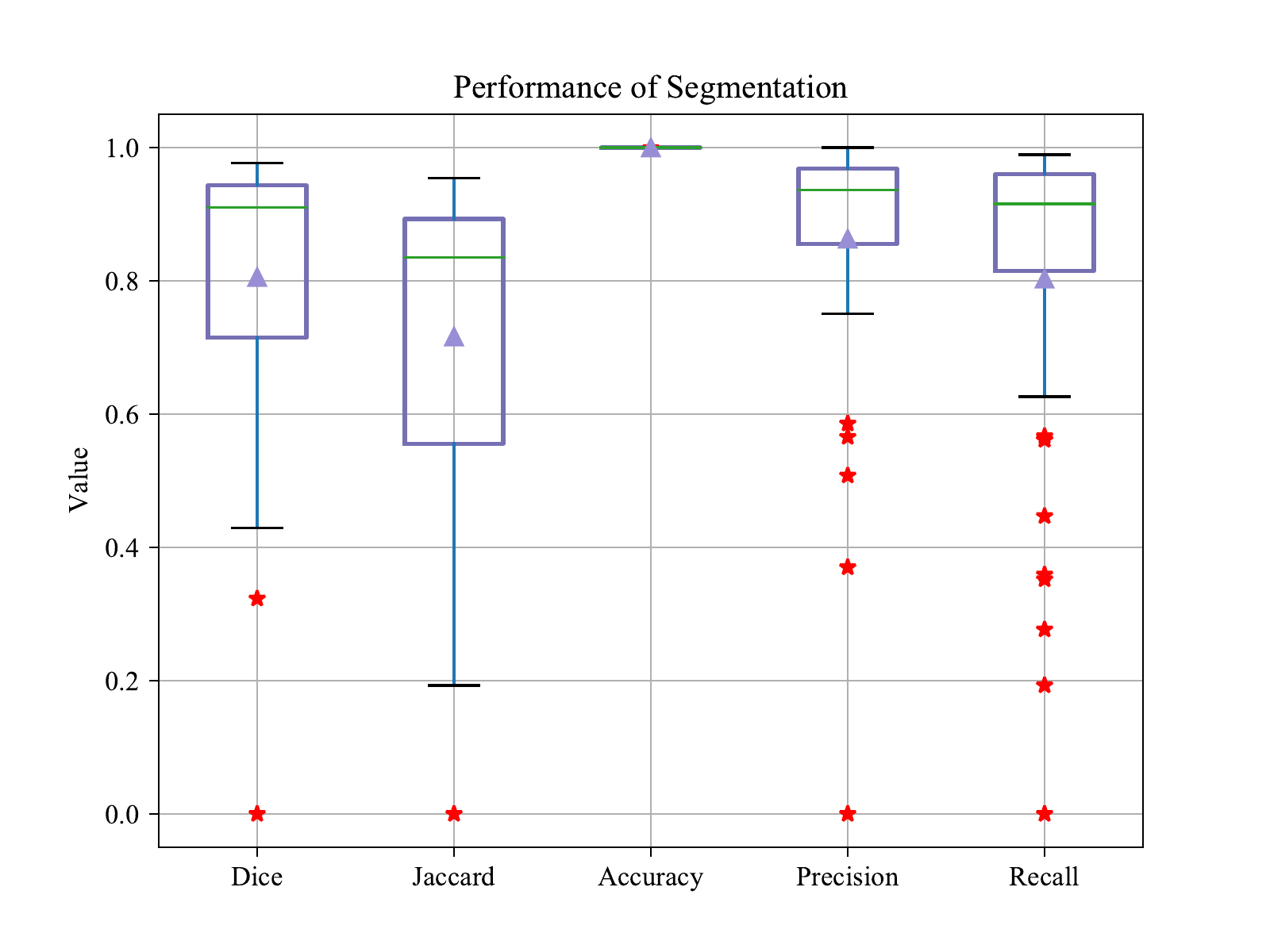}
        \caption{Tumor segmentation}
        \label{fig:tumor}
    \end{subfigure}
    \caption{Quantitative metrics}
\end{figure}

We evaluate all of the 42 test patient scans using the six quantitative metrics to evaluate our method objectively and comprehensively; Dice, Jaccard, Accuracy, Precision, Recall and Hausdorff are computed for both kidney and kidney tumor segmentation. To clearly observe the distribution of all the test patients, the indicators of Dice, Jaccard, Accuracy, Precision and Recall are gathered into two box plots, shown in Fig.~\ref{fig:kidney} and Fig.~\ref{fig:tumor} for kidney and kidney tumor segmentation, respectively. 

Our method is further elaborated from the basic 3D U-Net. We do not consider nor compare our approach to other complicated architecture modifications, as they are often only effective for specific cases or metrics. Therefore, in order to investigate the effectiveness of our strategies, we compare the performance of the basic 3D U-Net with our multi-scale supervised 3D U-Net. The implementation of the 3D U-Net is identical to our multi-scale supervised U-Net except for the three strategies defined in this paper. The comparison results are listed in Table~\ref{tab2} and Table~\ref{tab3}. These two tables present the average values of the six indicators and prove the improvement after using our three-fold enhancing strategies. The most notable differences are in terms of tumor segmentation.

\begin{table}
\centering
\caption{Comparison between the proposed MSS U-Net and classic 3D U-Net (kidney), the bold is better.}
\begin{tabular}{@{}lcc@{}}\toprule
\label{tab2}
\textbf{Metric}         & \textbf{MSS U-Net }       & \textbf{Classic 3D U-Net} \\ 
\midrule
Dice           & \textbf{\enspace 0.969}   & \enspace 0.962           \\
Jaccard        & \textbf{\enspace 0.941}   & \enspace 0.930           \\
Accuracy       & \textbf{\enspace 0.999}   & \enspace 0.999           \\
Precision      & \textbf{\enspace 0.971}   & \enspace 0.961           \\
Recall         & \textbf{\enspace 0.968}   & \enspace 0.965           \\
Hausdorff (mm) & \textbf{19.188}           & 38.945          \\ 
\bottomrule
\end{tabular}
\end{table}

\begin{table}
\centering
\caption{Comparison between the proposed MSS U-Net and classic 3D U-Net (tumor).}
\label{tab3}
\begin{tabular}{@{}lcc@{}}
\toprule
\textbf{Metric}         & \textbf{MSS U-Net }       & \textbf{Classic 3D U-Net} \\ 
\midrule
Dice           & \textbf{ 0.805}  & \enspace 0.781            \\
Jaccard        & \textbf{ 0.716}  & \enspace 0.699            \\
Accuracy       & \textbf{ 0.999}  & \enspace 0.999            \\
Precision      & \textbf{ 0.863}  & \enspace 0.841            \\
Recall         & { 0.802}         & \textbf{\enspace 0.810}   \\
Hausdorff (mm) & \textbf{33.469}  & 50.808                    \\ 
\bottomrule
\end{tabular}
\end{table}

After comparing with the basic U-Net architecture, we compare our approach with two recent methods listed in Table~\ref{tab4}. In this comparison, it should be noted that our proposed method processes raw-size CT images, which adds significant complexity to the sementation model when compared to the other two based on smaller ROIs. Nevertheless, our method still outperforms them in multiple metrics regarding both kidney and kidney tumor segmentation. 

\begin{table}
\centering
\caption{Comparison of Dice coefficient of the proposed MSS U-Net and state-of-the-art methods.}
\label{tab4}
\begin{tabular}{@{}lll@{}}
\toprule
\textbf{Method}                      & \textbf{Kidney}         & \textbf{Tumor}      \\ 
\midrule
2D\_PSPNET (ROI)~\cite{havaei2017brain}         & 0.902             & 0.638          \\
3D\_FCN\_PPM (ROI)~\cite{yang2018automatic}     & 0.927             & 0.802          \\
MSS U-Net (Raw images)                          & \textbf{0.969}    & \textbf{0.805} \\ 
\bottomrule
\end{tabular}
\end{table}

In addition, to compare our methods with other current state-of-the-art architectures and approaches, we participated in the KiTS19 challenge. During the challenge, we were able to perform a broader evaluation with the other state-of-the-art methods. The proposed network architecture, MSS U-Net, ranked in 7th position out of 106 teams~\cite{heller2019kits19} achieving a kidney and kidney tumor Dice of 0.974 and 0.818, respectively, measured using the data of the 90 test patients. This demonstrates both the applicability of the approach described in this paper for real data and its improved performance compared to previous methods relying only on the 3D-UNet architecture for medical image segmentation.

We list several top winners of KiTS19 challenge in Table~\ref{tab5}. Fabian et al.~\cite{isensee2019attempt}, the  first-ranked solution, proposed a residual 3D U-Net to enhance the segmentation performance. In this case, the authors modified part of the training data in order to gain a unique advantage. Xiaoshuai et al.~\cite{heller2019state} and Guangrui et al.~\cite{mu2019segmentation} constructed their methods using 3-stage and 2-stage segmentation respectively, which followed a different strategy compared to our motivation for having an end-to-end method. Finally, Andriy et al.~\cite{myronenko20193d} followed a more similar strategy to ours because they also focused on how to better train a basic 3D U-Net and employed the fashionable boundary-aware loss~\cite{hatamizadeh2019boundary}. However, they used a much bigger input (176×176×176). In general terms, we believe that our method can be further optimized by conducting more adaptive experiments due to the larger number of hyperparameters directed to training the 3D U-Net more efficiently.

\begin{table}
    \centering
    \caption{Dice of the proposed method and other algorithms in KiTS19 challenge. These results are obtained from the competition's test dataset. This test dataset is public, but the ground truth labels were not made public.}
    \label{tab5}
    \begin{tabular}{@{}lll@{}}
        \toprule
        \textbf{Method}                               & \textbf{Kidney} & \textbf{Tumor}  \\ 
        \midrule
        Fabian et al.~\cite{isensee2019attempt}, 1st place      & 0.974 & 0.851 \\
        Xiaoshuai et al.~\cite{heller2019state}, 2nd place      & 0.967 & 0.845 \\
        Guangrui et al.~\cite{mu2019segmentation}, 3rd place    & 0.973 & 0.832 \\
        Andriy et al.~\cite{myronenko20193d}, 9th place         & 0.974 & 0.810 \\
        MSS U-Net, 7th place                                    & 0.974 & 0.818 \\ 
        \bottomrule
    \end{tabular}
\end{table}

It is worth noting that, due to the limited size of the dataset, the results could vary significantly if only one tumor was not detected. This is particularly evident with particularly small tumors that can pass undetected simply because of sampling or data handling issues. Nonetheless, this is an important aspect that must be considered because detecting small tumors or lesions can be key in early disease detection. Thus, we believe that formal solutions need to be proposed to tackle this specific problem of small tumor segmentation.

\section{Discussion}

In recent years, deep learning based methods have accounted for the largest fraction of research papers in the field of medical image segmentation. A wide variety of networks with many new architectures have been proposed, with innovative properties and significantly superior results in multiple aspects when compared to more traditional methods. Nonetheless, among all these, and to the best of our knowledge, the original U-Net architecture is still able to achieve results comparable to the state-of-the-art and even outperform more recent architectures in certain aspects related to medical image segmentation~\cite{isensee2018nnu}. This has been further exemplified in this paper, as we have demonstrated the capabilities of the original architecture if more efficient training techniques are introduced as proved by our experimental results.

Based on the 3D U-Net architecture, we have designed and trained a six-layer network and proposed three effective strategies including both training techniques and data augmentation. First, we have employed multi-scale supervision to increase the probability of the network predicting low-resolution labels correctly from deeper layers. This concept can be better understood with an analogy of human behaviour: the action of first zooming out an image to label or identify coarse contours and then zoom into precisely label the image at a finer level. Second, in order to alleviate the inherent imbalance in samples and segmentation difficulty across organs, we have introduced the use of exponential logarithmic loss to induce the network into paying more attention to tumor samples and more difficult samples. Third, we have designed a connected-component based post processing method to remove the clearly mistaken voxels that have been identified by the network. 

\begin{figure*}
    \centering
    \fbox{\includegraphics[width=0.85\textwidth]{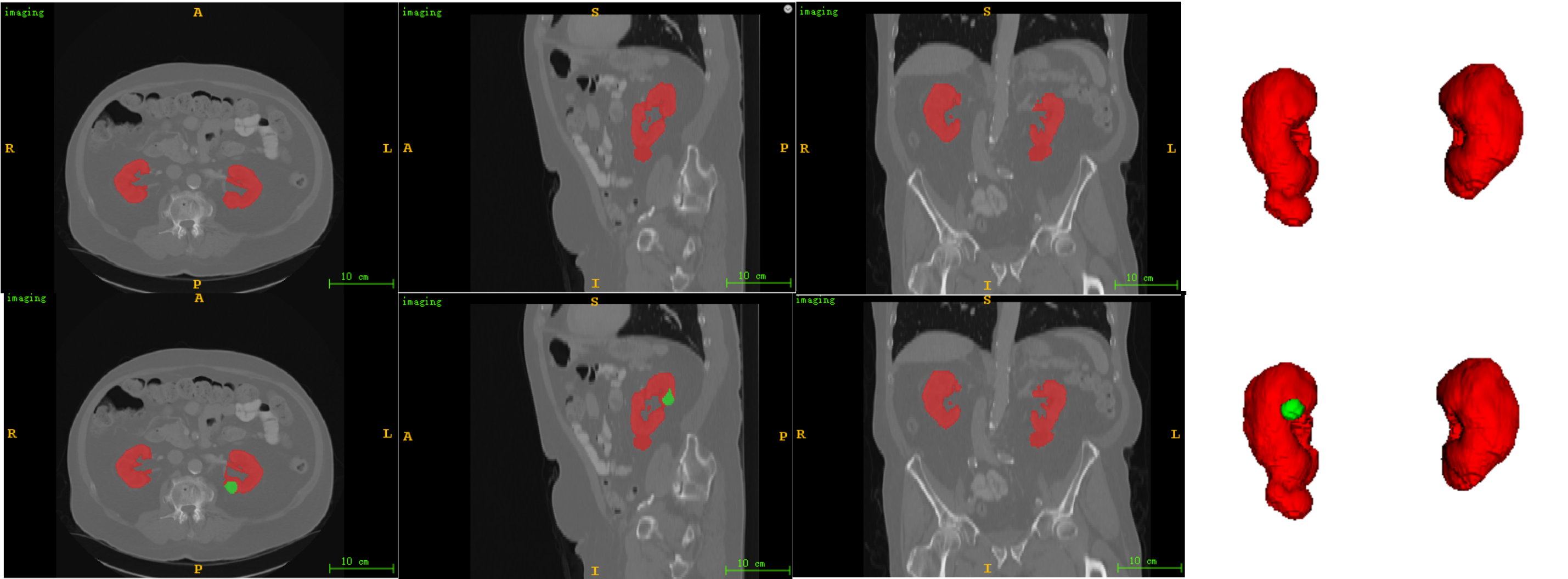}} \\ 
    \caption{Illustration of the worst-case segmentation from the proposed network. The top row shows the predicted segmentation while the bottom row shows the ground truth. The tumor, in green, is not identified by our network, which is only able to segment the kidney tissue instead, in red. The small size of the tumor in this case may have played a significant role in the segmentation failure.}
    \label{fig:worst}
\end{figure*}

The experiments that we have carried out and the comparisons with existing methods have shown the advantages of the proposed architecture and the effectiveness of the enhancement strategies over the original 3D U-Net. Nevertheless, several aspects remain challenging and require further investigation. In particular, from the segmentation statistics of the test patients, we have found that two patients had a low tumor Dice with one of them being as low as zero. The data corresponding to the latter patient is shown in Fig.~\ref{fig:worst}. In the figure, we can see that the prediction mask does not map the tumor, a consistent result with the Dice coefficient. The cause of this low Dice might be the particularly small size of the tumor in the CT images. We attribute this phenomenon to the fixed receptive field of classic convolution. Therefore, we will consider in our future work to employ deformable convolution as a potential solution to this problem~\cite{dai2017deformable}.  

In summary, the proposed architecture and chosen methodology in this paper are based on the assumption that the basic 3D U-Net architecture is capable of extracting sufficient features for segmentation. Therefore, we have directed our efforts towards the training procedures while discarding the complex architecture modifications with marginal effectiveness. At the time of finalizing the experiments reported in this paper, a modified U-Net (mU-Net) has been proposed by Seo \textit{et al.}~\cite{seo2019modified}. In their paper, the authors propose the utilization of residual path to the skip connections of the U-Net for the segmentation of livers and liver tumors. This paper shares a similar motivation with ours, aiming and innovative ways of mining more effective information from low-resolution features for accurate segmentation of medical images. The main difference from the architectural point of view is that Seo \textit{et al.} introduce additional blocks, while we focus on effective but simpler strategies to achieve the same goal.

\section{Conclusion}

In this paper, we have proposed an end-to-end multi-scale supervised 3D U-Net to simultaneously segment kidneys and kidney tumors from raw-scale computed tomography images. Extending the original 3D U-Net architecture, we have combined a multi-scale supervision approach with exponential logarithmic loss. This has enabled further optimization of the U-Net architecture, extending its possibilities, and hence obtaining better performance. Compared with a current trend in deep neural networks with complex architectures and multiple different submodules, we have taken a more generalized approach yet obtained results comparable to the state of the art. A simpler architecture has the advantage of higher reproducibility and wider generalization of results, in contrast with the potentially highly inflated models and poor reproducibility of more complex architectures.

In general, we have directed our efforts towards a more efficient training of the original 3D U-Net architecture by incorporating the multi-scale supervision and the exponential logarithmic loss. We have demonstrated the advantages of this approach with our experiments and comparisons with the state-of-the-art. While our architecture can be outperformed by others in specific metrics on the KiTS19 dataset, we have argued that having a simpler architecture still leaves room for further optimization and discussed the advantages from the point of view of applicability and extendability.

Finally, the code leading to this work has been made public through a GitHub repository\footnote{https://github.com/LINGYUNFDU/MSSU-Net}, with the code that has been used on the KiTS19 dataset. In the future work, we expect to extend the application of our architecture towards segmentation of other organs and modalities, such as magnetic resonance imaging (MRI) or positron-emission tomography (PET).


\nocite{*}
\bibliographystyle{unsrt}
\bibliography{main.bib}

\end{document}